\documentclass{iopjournal}
\usepackage{amsmath,amssymb}
\usepackage{bm}
\usepackage{slashed}

\begin{document}

\articletype{Letter} %	 e.g. Paper, Letter, Topical Review...

% \title{Realizing Symmetry-Protected Topological States through Polyakov's Confinement Mechanism in Triangular Antiferromagnets}
\title{Emergent Symmetry-Protected Topological Phases via Polyakov Confinement in Quantum Spin Systems}

\author{Li-Wei He$^1$\orcid{0009-0001-9897-545X}, Shun-Li Yu$^{1,2,3}$\orcid{0000-0001-7202-4851} and Jian-Xin Li$^{1,2,3,*}$\orcid{0009-0007-3901-8119}}

\affil{$^1$National Laboratory of Solid State Microstructures and Department of Physics, Nanjing University, Nanjing 210093, China}

\affil{$^2$Collaborative Innovation Center of Advanced Microstructures, Nanjing University, Nanjing 210093, China}

\affil{$^3$Jiangsu Key Laboratory of Quantum Information Science and Technology, Nanjing University, Suzhou 215163, China}

\affil{$^*$Author to whom any correspondence should be addressed.}

\email{slyu@nju.edu.cn and jxli@nju.edu.cn}


% Supplementary material for this article is available online

\begin{abstract}
A main theme in modern condensed matter physics is the emergence of fractionalized excitations and gauge structures from quantum spin systems. However, understanding how these exotic degrees of freedom reconfine into new phases of matter remains a fundamental challenge. In this work, we demonstrate that Polyakov’s confinement mechanism—a foundation‌ of compact gauge theory—can serve as a dynamical engine to transform a Dirac spin liquid into a symmetry-protected topological (SPT) state. Starting with a Dirac spin liquid with spin-dependent gauge fluxes, we show that the single-occupancy constraint inherent to the physical Hilbert space triggers monopole condensation, dynamically confining bulk spinons while preserving gapless edge modes—realizing a spinon analog of the quantum spin Hall effect. Using a large-scale variational Monte Carlo simulation on a triangular antiferromagnet with an additional Dzyaloshinskii–Moriya interaction, we provide microscopic evidence for this confined SPT phase, including a characteristic area law for the Wilson loop and vanishing topological entanglement entropy. Furthermore, we identify a measurable spin-pump response under magnetic fields, which directly encodes the Berry curvature of the parent Dirac cones. Our results reveal a previously unexplored pathway to SPT physics, bridging the fields of gauge theory, quantum magnetism, and topological matter.
\end{abstract}

\section{Introduction}
Quantum spin liquids (QSLs)~\cite{mat.res.bull.8.153,RepProgPhys.80.016502,RevModPhys.89.025003,npj.quan.mater.4.12,sci.367.eaay0668,RevModPhys.97.045003}, which lack magnetic order even at zero temperature and are characterized by fractional excitations and long-range quantum entanglement, represent a distinct paradigm beyond the Landau-Ginzburg framework of symmetry breaking and have garnered significant interest in condensed matter physics. The low-energy physics of these exotic quantum states can be effectively described by quantum field theories (QFTs). For example, the gapped chiral spin liquid (CSL) with semionic topological excitations~\cite{PhysRevB.39.11413},  realized in Heisenberg spin models on frustrated (triangular and kagome) magnets~\cite{PhysRevB.94.075131,PhysRevLett.127.087201,PhysRevB.108.245102,nat.commun.5.5137,PhysRevLett.115.267209,PhysRevB.91.041124}, is fully described by the $U(1)$ Chern-Simons theory with the concise $K$-matrix $K = (2)$~\cite{Natl.Sci.Rev.3.68}. This state is also known as the $\nu = 1/2$ Laughlin state, a foundational instance of topological order.

According to the classification theory of projective symmetry groups~\cite{PhysRevB.93.094437}, this CSL can be viewed as a parity-time-breaking instability of the gapless Dirac spin liquid (DSL) in frustrated spin models.This algebraic DSL realizes the well-known quantum electrodynamics in $2+1$ dimensions (QED$_3$ with $N_f = 4$ flavors) as a low-energy gauge QFT description~\cite{nc.commun.10.4254,PhysRevX.10.011033,PhysRevX.14.021010}. The gap of the descendant CSL arises from massless Dirac spinons coupled to an emergent $U(1)$ gauge field with the same chirality. By contrast, if Dirac spinons of opposite spin couple to gauge fields of opposite chirality, a scenario reminiscent of double-semion topological order built from chiral-conjugate topological QFTs~\cite{PhysRevB.71.045110}, the resulting descendant state will preserve time-reversal symmetry while exhibiting nontrivial boundary physics. The emergent low-energy field theory and the fate of gauge confinement in such a time-reversal–symmetric descendant therefore constitute an important open problem.

This scenario presents a profound theoretical challenge: while research on QSLs has primarily focused on intrinsic topological order with long-range entanglement, their symmetric descendants may fall into distinct categories of quantum matter, such as symmetry-protected topological (SPT) phases, which lack anyonic bulk excitations but host robust, anomalous
boundary modes protected by global symmetries. Bridging the strongly correlated gauge fluctuations inherent in a QSL with the short-range entangled vacuum of an SPT phase is essential for developing a comprehensive classification of topological quantum states.

In this article, we establish that coupling Dirac spinons to a spin-dependent gauge field leads to a gapped, time-reversal-symmetric SPT state via Polyakov's confinement mechanism~\cite{PhysLettB.59.82}, a cornerstone of $2+1$D gauge theory. This mechanism, where the proliferation of space-time instantons (monopoles) confines fractionalized charges, is revealed here in a novel form. We show that when the gauge dynamics are linked to the spin sector, confinement does not destroy the topological properties. Instead, it acts as a dynamical filter, removing bulk gauge degrees of freedom while distilling a non-trivial SPT state. Our effective field theory illustrates that monopole condensation confines bulk spinons yet preserves protected edge modes, exhibiting a spinon analog of the quantum spin Hall effect. This theoretical framework is supported by numerical evidence from variational Monte Carlo (VMC) calculations, indicating that this state can be realized in a spin-1/2 triangular antiferromagnetic Heisenberg model with Dzyaloshinskii-Moriya (DM) interaction. Finally, we discuss key observable diagnostics from spin pumping under a magnetic field, providing a concrete experimental signature of this state. Our results demonstrate that Polyakov's confinement, far from being a prohibitive barrier for topological phases, can serve as a sophisticated dynamical engine for the emergence of symmetry-protected topology.

\section{Effective field theory description}\label{sec2}

\begin{figure}
	\centering
	\includegraphics[width=0.8\linewidth]{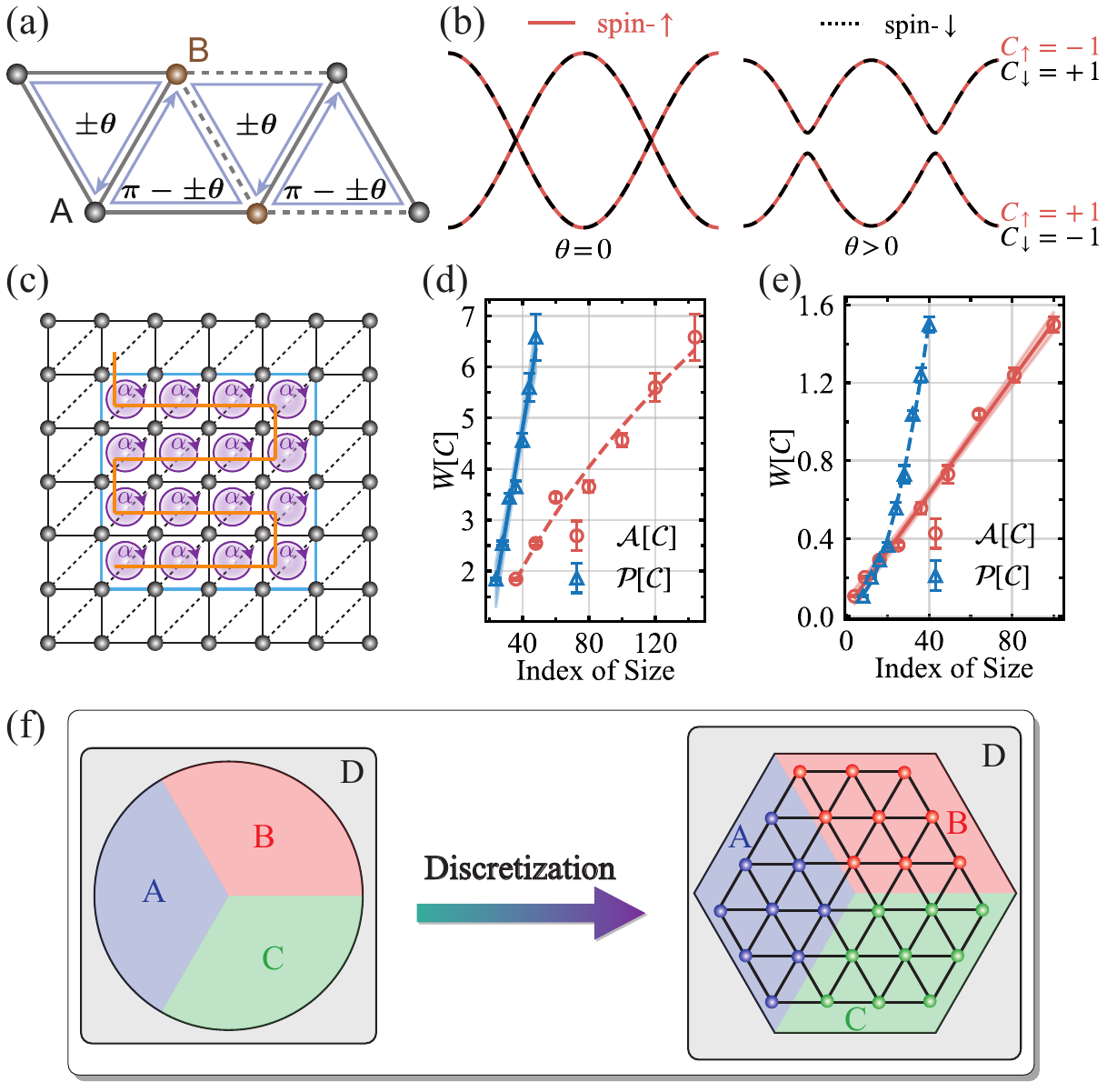}
	\caption{(a) The details of mean-field Ans\"atz, where the flux through down-pointing (up-pointing) triangles is $\pm \theta$ ($\pi \pm \theta$). The sign ``$\pm$" indicates that  spinons with opposite spins couple to opposite gauge fluxes: $\mathrm{e}^{i\theta_{\sigma}}c_{i,\sigma}^{\dagger}c_{j,\sigma}$ on the NN bond $\langle ij \rangle$, with $\theta_{\uparrow (\downarrow)} = +(-)\theta/3$ as the Peierls phase. This reduces to a DSL , depicted by the solid orange line in the phase diagram in Fig.~\ref{fig:phase_diagram}, when the gauge flux $\theta = 0$. A finite $\theta > 0$ opens gaps at Dirac cones, resulting in a topologically nontrivial Chern number $C_{\uparrow, \downarrow} = \pm 1$. The mean-field dispersion for two spin flavors in these cases is shown in (b). (c), Schematic illustration of flux insertion by a string operator, represented by the solid orange line on the distorted triangular lattice, where $\alpha$ denotes the gauge flux through an elementary plaquette. (d) and (e), $W[\mathcal{C}]$ as a function of the enclosed area ($\mathcal{A[C]}$) and perimeter ($\mathcal{P[C]}$) of the loop $\mathcal{C}$ (blue rings in (c)) for the DSL [$\theta =0$, system size $26 \times 26$ and period-period (P-P) boundary condition (BC)] and SPT state ($\theta=\pi/2$, chosen to maximize the mean-field gap, system size $24 \times 24$, P-P BC), respectively. (f), Application of the Kitaev–Preskill construction (left) to a discretized triangular lattice (right) for extracting the topological entanglement entropy. The simply connected disk is divided into three equal-area, simply connected subsystems A, B, and C (indicated by different colors), with D as the complementary region.}
	\label{fig:ek_dsl_spt_loop_subsys}
\end{figure}

Our starting point is the DSL, suggested in the triangular $J_1$-$J_2$ Heisenberg model~\cite{PhysRevX.14.021010,PhysRevB.93.144411} and kagome antiferromagnets with only $J_1$ term~\cite{PhysRevLett.98.117205,PhysRevX.7.031020,sciadv.aat5535,PhysRevLett.133.096501}. In the mean-field approximation with the Ansätz shown in Fig.~\ref{fig:ek_dsl_spt_loop_subsys}(a)  ~\cite{PhysRevB.93.144411,PhysRevLett.98.117205,PhysRevLett.133.096501}, the two Dirac cones for spinons are located at the momenta $\pm \boldsymbol{b}_{2}/4$ [see the left panel of \ref{fig:ek_dsl_spt_loop_subsys}b]. Near any Dirac point $\boldsymbol{K}$, the low-energy effective Hamiltonian is given by
\begin{equation}
    H_{0} = v\sum_{\sigma} \int \frac{d^2 \bm{q}}{{2\pi}^2} \psi^{\dagger}_{\sigma}(\boldsymbol{q})(q_x\tau^{x} + q_y\tau^{y})\psi_{\sigma}(\boldsymbol{q}),
\end{equation}
where the two-component spinor is $\psi_{\sigma}(\bm{q})=[c_{A,\sigma}(\bm{K+q}),c_{B,\sigma}(\bm{K+q})]^T$, with $A$ and $B$ being the sublattice indices. The single-occupation condition $c_{i,\uparrow}^{\dagger}c_{i,\uparrow} + c_{i,\downarrow}^{\dagger}c_{i,\downarrow} = 1$ can be enforced by including a gauge field $a_{\mu}$, leading to the corresponding Lagrangian for flavor $\sigma$:
\begin{equation}
    \mathcal{L}_{0} = \bar{\psi}_{\sigma}[i\gamma^{\mu}(\partial_{\mu} - i a_{\mu})]\psi_{\sigma} =\bar{\psi}_{\sigma}(i\slashed{D})\psi_{\sigma},
\end{equation}
where Feynman slash notation $\slashed{D} = \gamma^{\mu}D_{\mu}$ is used for simplicity. To generate the gauge field with opposite chirality, we introduce a finite Peierls phase $\theta_{\sigma} = \theta/3$ [see Fig.~\ref{fig:ek_dsl_spt_loop_subsys}(a)], which results in a mass term $\mathcal{L}_{mass} = -m_{\sigma}\bar{\psi}_{\sigma}\psi_{\sigma}$ with $m_{\uparrow} = -m_{\downarrow} = m$. Consequently, the Lagrangian can be expressed as
\begin{equation}
    \mathcal{L}_{\sigma} = \bar{\psi}_{\sigma}(i\slashed{D} - m_{\sigma})\psi_{\sigma}.
\end{equation}
As shown in the right panel of Fig.~\ref{fig:ek_dsl_spt_loop_subsys}(b), the Dirac cones are gapped in this case. By performing algebraic manipulations [see Supplementary material (SM) for details~\cite{supplement}], we obtain the effective Chern-Simons Lagrangian density:
\begin{equation}
\begin{aligned}
    \mathcal{L}_{\mathrm{CS}} &= \frac{1}{4\pi} \epsilon^{\mu \nu \rho} (a_{\mu}^{\uparrow} \partial_{\nu} a_{\rho}^{\uparrow} - a_{\mu}^{\downarrow} \partial_{\nu} a_{\rho}^{\downarrow})\\
    &= \frac{1}{4\pi} \epsilon^{\mu \nu \rho} \bm{a}_{\mu}^{T} K \partial_{\nu} \bm{a}_{\rho},
\end{aligned}
\end{equation}
where $\bm{a}_{\rho} = (a_{\rho}^{\uparrow}, a_{\rho}^{\downarrow})^{T}$, $K = \begin{pmatrix}
        1 & 0\\
        0 & -1
    \end{pmatrix}$,
and $\epsilon^{\mu \nu \rho}$ is the antisymmetric Levi-Civita tensor. The $K$ matrix clearly reveals that the spinons with opposite spin produce opposite Hall responses [or opposite Chern numbers, $C_{\uparrow,\downarrow} = \pm 1$, as shown in right panel of Fig.~\ref{fig:ek_dsl_spt_loop_subsys}(b)] and have a chiral central charge $c = 0$. This demonstrates that the global time-reversal symmetry is preserved, despite being individually broken for each spin component. Additionally, the ground-state degeneracy (GSD) is $n_{g} =1$ (see SM for details~\cite{supplement}), meaning this Chern-Simons theory describes a SPT state. According to the group-cohomology classification of bosonic SPT phases~\cite{PhysRevB.89.035147}, it is characterized by a nontrivial element in $\mathcal{H}^{3}\bigl(U(1) \rtimes \mathbb{Z}_{2}^{\mathcal{T}}, U_{\mathcal{T}}(1)\bigr)$. Hence, this SPT state exhibits the same topological structure as the quantum spin Hall state~\cite{PhysRevLett.96.106802}, but arises from spinons rather than electrons.

Going beyond the mean-field approximation, we must take into account the single-occupancy constraint. In the projective construction theory, 
this constraint can be enforced by a Gutzwiller operator $P_{G}$, which restores the underlying lattice gauge structure, causing the emergent gauge field to become compact. As a result, $a_{\nu}$ is now defined modulo $2\pi$~\cite{PhysRevD.10.2445,PhysRevD.10.3376,PhysLettB.59.82}, and monopole (instanton) configurations that change the total gauge flux by integer multiples of $2\pi$ are allowed. Therefore, monopole events, which were absent in the non-compact mean-field description, must be taken into account in the projected (physical) theory. With the Chern–Simons term $\mathcal{L}_{\mathrm{CS}}$ canceled ($\mathrm{Tr}(K) = 0$), the effective theory contains only the compact Maxwell term and monopole contributions. The effective low-energy Lagrangian in Euclidean spacetime is then given by
\begin{equation}
    \mathcal{L} = \frac{1}{4g^2}F_{\mu \nu} F^{\mu \nu}  + \mathcal{L}_{\mathrm{mono}},
\end{equation}
where the first term is the Maxwell term, arising from the parity-even component of the polarization tensor, with coupling factor $g$ and field strength tensor $F_{\mu \nu} = \partial_{\mu}a_{\nu} - \partial_{\nu}a_{\mu}$.  The final term corresponds to the monopole instanton in the compact $U(1)$ gauge field.

To make the effects of monopole events more explicit and to handle the compact gauge field more conveniently, we introduce a dual scalar field $\chi$ via a Hubbard–Stratonovich transformation, where $F_{\mu \nu} \leftrightarrow \epsilon^{\mu \nu \rho} \partial^{\rho} \chi$ and $\mathcal{L}_{\mathrm{mono}} \sim -2\xi \cos \chi$~\cite{PhysRevD.10.3376,PhysLettB.59.82}, with $\xi$ being the instanton fugacity. The dual Lagrangian then reads
\begin{equation}
    \mathcal{L}_{\mathrm{dual}} = \frac{1}{2g^2}(\partial_{\mu} \chi)^2 - 2 \xi \cos \chi.
\end{equation}
To analyze the low-energy behavior, we expand the dual scalar field $\chi$ around the minimum of the cosine potential, yielding
\begin{equation}
    \mathcal{L}_{\mathrm{dual}} \to  \frac{1}{2g^2}(\partial_{\mu} \delta \chi)^2 + \xi (\delta \chi)^{2}.
\end{equation}
This gives rise to a dynamically generated mass $m = \sqrt{2 \xi g^2}$ for the dual photon, originating from monopole proliferation in a compact $U(1)$ gauge theory in $2+1$ dimensions. Consequently, a linear potential emerges between any pair of gauge charges carried by spinons, indicating that spinons are confined in the bulk. As a result, fractional excitations exist only at the edges. This analysis of spinon confinement serves as an analogy to the fundamental framework proposed by Polyakov for understanding quark confinement~\cite{PhysLettB.59.82}.

\section{Diagnostics  of confinement via  Wilson Loop  and  topological entanglement entropy}

%A defining hallmark of Polyakov confinement is the transition from a perimeter-law to an area-law scaling of the Wilson loop operator—a diagnostic that directly probes the fate of gauge charges in the bulk. To verify that our time-reversal-symmetric state indeed resides in the confined phase predicted by the field theory, we employ a lattice implementation of this diagnostic using variational Monte Carlo (VMC) calculations.

To substantiate the spinon confinement predicted by the field theory, we directly probe the bulk characteristic using two complementary numerical diagnostics.  First, we compute 
the correlation function introduced by Wilson~\cite{PhysLettB.59.82,PhysRevD.10.2445}:
\begin{equation}
    e^{-W[\mathcal{C}]} = \langle e^{\oint_{\mathcal{C}} A_{\mu} dx_{\mu}} \rangle,
\end{equation}
where $\mathcal{C}$ is a closed loop. The Wilson loop serves as the definitive diagnostic of confinement: in a deconfined phase, it follows a perimeter law, $W[\mathcal{C} ] \sim \mathcal{P[C]}$,  whereas in a confined phase, it obeys an area law,  $W[\mathcal{C}] \sim \mathcal{A[C]}$
%The defining feature of the charge confinement (deconfinement) is that  $W[\mathcal{C}]$ is proportional to the exponential of the loop's area $\mathcal{A[C]}$ (perimeter $\mathcal{P[C]}$). } 
In the lattice system, to introduce gauge flux into the loop $\mathcal{C}$, we define a discrete gauge string operator $\hat{\mathcal{S}}$ that intersects a sequence of $N$ real-space bonds, depicted as solid links in Fig.~\ref{fig:ek_dsl_spt_loop_subsys}(c). Along the extent of the string, the gauge phase $\phi_j$ for the $j$-th bond increases linearly (Landau gauge):
\begin{equation}
    \phi_j = j \alpha, \quad (j = 1, 2, \dots, N)
\end{equation}
where $\alpha = 2\pi / N$ is the elementary flux increment, ensuring the total phase winding satisfies $N\alpha = 2\pi$. This configuration effectively traps a $2\pi$ gauge flux within $\mathcal{C}$, providing a direct probe of gauge confinement, analogous to Polyakov's original formulation~\cite{PhysLettB.59.82,PhysRevD.10.2445}. The quantity $W[\mathcal{C}] = -\ln |\langle \hat{\mathcal{S}} \rangle|$ can be calculated using the VMC method with projective wave functions.  The results for this quantity as a function of the enclosed area ($\mathcal{A[C]}$) and perimeter ($\mathcal{P[C]}$) of the loop $\mathcal{C}$ for the DSL ($\theta =0$) and SPT state ($\theta=\pi/2$) are shown in Fig.~\ref{fig:ek_dsl_spt_loop_subsys}(d) and (e), respectively. It shows clearly that $W[\mathcal{C}] \sim \mathcal{P[C]}$ for the deconfined DSL, while $W[\mathcal{C}] \sim \mathcal{A[C]}$ for the SPT state with bulk confinement. Although our lattice implementation differs in microscopic details from the continuum formulation, we believe the emergence of this area law serves as a diagnostic for the confinement mechanism envisioned by Polyakov.

%Second, the system with Polyakov's confinement will exhibit only short-range entanglement. So, a key distinction between this SPT state and gapped topological QSLs is that the topological entanglement entropy (TEE) of this state is zero. To characterize the entanglement structure of this SPT state, we compute the TEE. 
Second, Polyakov confinement not only gaps the bulk but also ensures that the ground state exhibits only short-range entanglement. Consequently, in contrast to gapped topological QSLs, this SPT phase is characterized by a vanishing topological entanglement entropy (TEE). To confirm this and quantitatively characterize the entanglement structure, we extract the TEE using the Kitaev-Preskill construction~\cite{PhysRevLett.96.110404}.
Without loss of generality, we can reparametrize the Peierls phase with the nearest-neighbor (NN) hopping term as:
\begin{equation}
\mathrm{e}^{i\theta_{\sigma}}c_{i,\sigma}^{\dagger}c_{j,\sigma} \to (1 + i\sigma t_{z})c_{i,\sigma}^{\dagger}c_{j,\sigma},
\label{tz-eq}
\end{equation}
and partition the system into three subsystems A, B, and C, arranged with $C_{3}$
rotational symmetry adapted to the triangular lattice.
%Imitating the Kitaev-Preskill scheme~\cite{PhysRevLett.96.110404}, we design a separation of three subsystems A, B, and C with $C_{3}$ rotational symmetry, considering the triangular geometry. 
A well-defined factorization of the Hilbert space requires that each degree of freedom (i.e., each lattice site) be assigned to one and only one subsystem; hence, the three subsystems A, B, and C must be mutually disjoint ($\mathrm{A} \cap \mathrm{B} = \mathrm{B}\cap \mathrm{C} = \mathrm{A}\cap \mathrm{C} = \varnothing$), as shown in Fig.~\ref{fig:ek_dsl_spt_loop_subsys}(f). With this separation, the TEE $\gamma = -(S_{\mathrm{A}} + S_{\mathrm{B}} + S_{\mathrm{C}} - S_{\mathrm{AB}} - S_{\mathrm{AC}} - S_{\mathrm{BC}} + S_{\mathrm{ABC}})$ can be reduced to $\gamma = -(3S_{\mathrm{A}} - 3S_{\mathrm{AB}} + S_{\mathrm{ABC}})$, where $S$ is the two-order Renyi entropy of the subsystem~\cite{PhysRevLett.103.261601}. Our numerical simulations confirm the expected symmetry relations $S_{\mathrm{A}} = S_{\mathrm{B}} = S_{\mathrm{C}}$ and $S_{\mathrm{AB}} = S_{\mathrm{AC}} = S_{\mathrm{BC}}$ within numerical error. We obtain the TEE $\gamma \sim 0 = \ln 1 = \ln D_{q}$ ($D_{q} = 1$ is quantum dimension of this state), as listed in Tab.~\ref{tab:tee}, indicating that this projective many-body state exhibits short-range entanglement. This behavior is fully consistent with the nature of an SPT state.

\begin{table}
    \centering
    \caption{TEE as a function of $t_{z}$ in equation~(\ref{tz-eq}) and the subsystem size $N_{\mathrm{A}}$. The total system size is $N = 12 \times 12 = 144$.}
    \begin{tabular*}{\linewidth}{@{\extracolsep\fill}l c c}
        \hline \hline
        $t_{z}$ & $N_{\mathrm{A}}$ & $\gamma$ \\
        \hline
        0.05 & $4 \times 4$  &  0.015 $\pm$ 0.188\\
        0.1 & $4 \times 4$  &  0.000 $\pm$ 0.155\\
        0.2 & $4 \times 4$  &  -0.090 $\pm$ 0.087\\
        \hline \hline
    \end{tabular*}
    \label{tab:tee}
\end{table}

\section{Model realization}

To concretely realize the proposed SPT phase, we consider the following spin model on a triangular lattice.
\begin{equation}
    \label{eq:model}
    H = J_{1}\sum_{\langle ij \rangle} \boldsymbol{S}_{i} \cdot \boldsymbol{S}_{j} + J_{2}\sum_{\langle \langle ij \rangle \rangle} \boldsymbol{S}_{i} \cdot \boldsymbol{S}_{j} + D\sum_{\langle ij\rangle} \boldsymbol{n}_{z} \cdot \boldsymbol{S}_{i} \times \boldsymbol{S}_{j},
\end{equation}
where the $J_{1,2}$ terms represent the nearest-neighbor (NN) and next-nearest-neighbor (NNN) antiferromagnetic Heisenberg couplings, respectively. The last term is the DM interaction, with $\boldsymbol{n}_{z}$ being the unit vector along the $z$-direction. We then derive the mean-field (mf) Hamiltonian:
\begin{align}
    H_{\mathrm{mf}} &= \bigl[\sum_{\langle ij \rangle} \psi^{\dagger}_{i} (t_{1,ij} + it_{z,ij}\sigma^{z}) \psi_{j} + \sum_{\langle \langle ij \rangle \rangle} t_{2,ij} \psi^{\dagger}_{i} \psi_{j} + \mathrm{H.c.} \bigr] \nonumber \\
    &+ \sum_{i} \psi^{\dagger}_{i} (\boldsymbol{M}_{i} \cdot \frac{\boldsymbol{\sigma}}{2}  + \lambda) \psi_{i},
\label{eq:hmf}
\end{align}
see SM~\cite{supplement} for details.
% where $t_{1,ij}$, $t_{z,ij}$ and $t_{2,ij}$ are hopping terms. $\boldsymbol{M}_{i} = M \bigl(\cos(\boldsymbol{Q} \cdot \boldsymbol{r}_{i}), \sin(\boldsymbol{Q} \cdot \boldsymbol{r}_{i}), 0\bigr)$ is the static background field describing the classical magnetic order, and $\lambda$ is the Lagrange multiplier. Thus, the parameters space is $p = (t_{1,ij}, t_{z,ij}, t_{2,ij}, M, \lambda)$, which are determined through VMC calculations.

\begin{figure}
    \centering
    \includegraphics[width=0.8\linewidth]{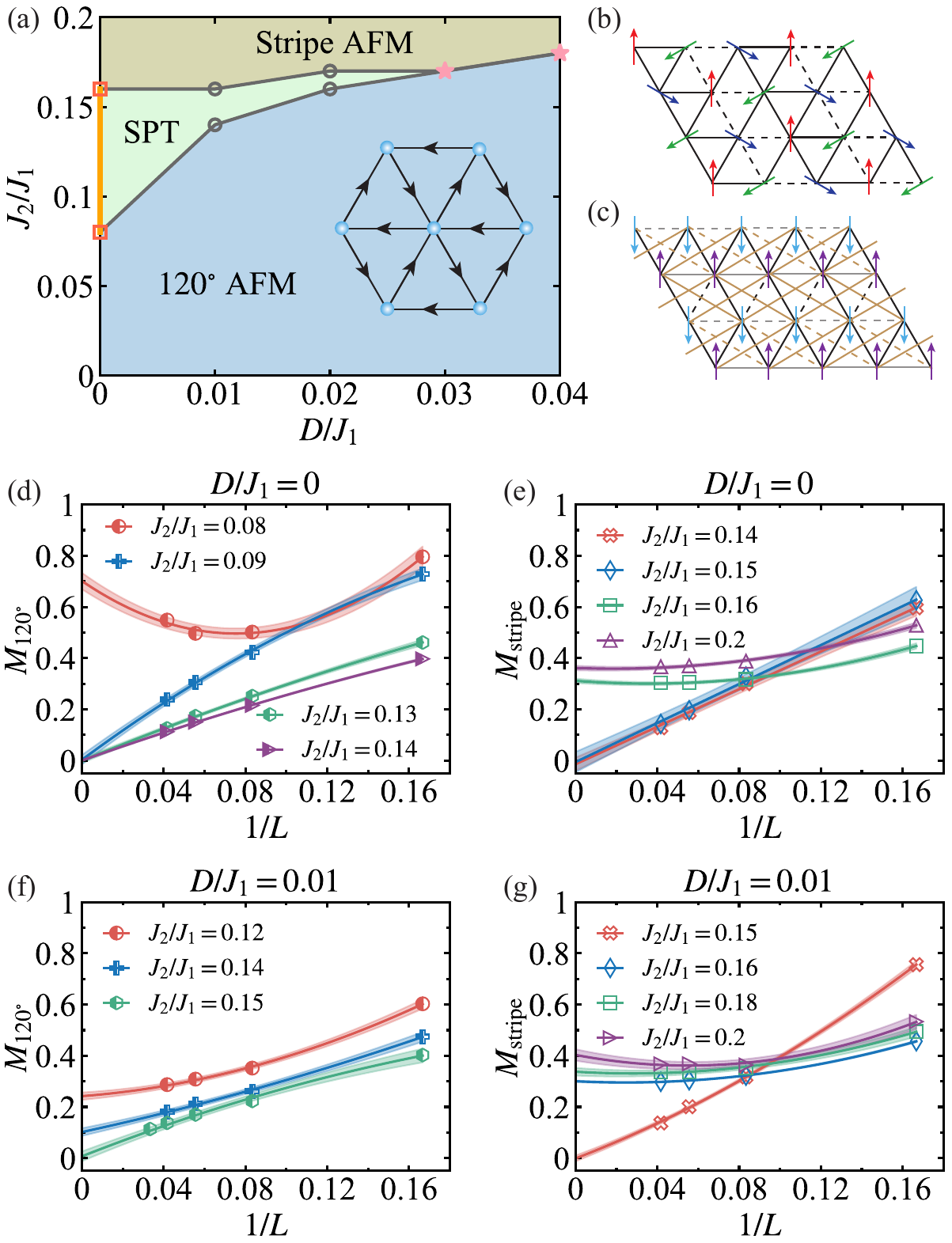}
    \caption{(a) Phase diagram of the $J_{1}$-$J_{2}$ antiferromagnetic model with DM interaction on a triangular lattice. The inset illustrates the pattern of the DM interaction, where the vector $\boldsymbol{D}$ indicates the spin direction along the $+z$-axis. (b) and (c), The mean-field Ans\"atze of the $120^{\circ}$ and strip antiferromagnetic orders, respectively. Different spinon hopping terms are highlighted in various colors, with dashed lines indicating terms that include an additional negative sign. The optimized variational magnetic moments $M$ for the $120^{\circ}$ [(d) and (f)] and stripe [(e) and (g)] orders are shown as a function of $1/L$ (where the system size is $L \times L$) for different DM and $J_2$ spin interactions. The solid lines in various colors represent quadratic extrapolations, and the shaded areas indicate the confidence intervals from the fitting.}
    \label{fig:phase_diagram}
\end{figure}

When the DM interaction $D = 0$, we identify a DSL phase in the range of $0.08 < J_{2}/J_{1} < 0.16$, which aligns quantitatively with earlier VMC calculations~\cite{PhysRevB.93.144411}. This is depicted by the orange solid line between the $120^{\circ}$ and stripe antiferromagnetic phases in the phase diagram Fig.~\ref{fig:phase_diagram}(a). In the $120^{\circ}$ antiferromagnetic phase for $J_{2}/J_{1} < 0.08$, with a pitch vector $\boldsymbol{Q} = (1/3, 1/3)$ in units of the primitive reciprocal lattice vectors, there are alternative fluxes $0$ and $\pi$ through the triangles, as shown in Fig.~\ref{fig:phase_diagram}(b). When the NNN Heisenberg term satisfies $J_{2}/J_{1} \geq 0.16$, a collinear stripe antiferromagnetic order emerges with an ordering vector $\boldsymbol{Q} = (0, 1/2)$. In this phase, the spinon hopping terms [see Fig.~\ref{fig:phase_diagram}(c) for the details] undergo a nematic instability with a finite NNN hopping term $t_{2,ij}$~\cite{PhysRevB.93.165113}. The phase transitions are determined by the extrapolations, as illustrated in Fig.~\ref{fig:phase_diagram}(d) and (e).

When a weak DM interaction is introduced, the optimized spin-triplet hopping term $t_{z,ij}$ in equation~(\ref{eq:hmf}) becomes finite and purely imaginary, according to our VMC calculations. As $J_{2}$ increases, the $120^{\circ}$ magnetic order is gradually suppressed, transitioning into a disordered phase. Compared to the scenario without DM interaction, the critical $J_{2}$ is larger [see Fig. \ref{fig:phase_diagram}(a)]. This indicates that the DM interaction, like the antiferromagnetic $J_{1}$ term~\cite{PhysRevLett.127.127205}, also favors the $120^{\circ}$ order. In the disordered phase, the optimized $t_{2,ij}$ vanishes, and the remaining hopping parameters form the Ans\"atz of the SPT phase with time-reversal symmetry, as discussed earlier. For large $J_{2}$, the stripe antiferromagnetic order is energetically favored. The $t_{z,ij}$ term breaks rotational symmetry and exhibits nematic characteristics, similar to other hopping terms in the absence of DM interaction. We determine the phase boundary through extrapolations in the presence of finite DM interaction, as shown in Fig.~\ref{fig:phase_diagram}(f) and (g) for $D/J_{1} =0.01$. Additional results for $D/J_{1} > 0.01$ are provided in the SM~\cite{supplement}. When $D \geq 0.03$, the phase transition between the $120^{\circ}$ and stripe antiferromagnetic orders is first-order, indicating the absence of a deconfined quantum critical point [see SM for details~\cite{supplement}]. Thus, the interplay between the antiferromagnetic $J_{2}$ term and DM interaction in the triangular lattice antiferromagnet can indeed realize the SPT state.

We further uncover a second SPT phase within the parameter window $0.13 \le J_2/J_1 \le 0.15$ when the DM interaction parameter $D$ is negative ($D/J_1=-0.01$) (see SM for details~\cite{supplement}). These two SPT states are orthogonal, possessing opposite spin Chern numbers and both exhibiting protected edge spin responses. Consequently, the DM interaction $D$ acts as a tuning parameter that drives a continuous phase transition from one SPT state to another through a deconfined quantum critical point at $D=0$, where the system becomes a DSL.

\begin{figure}
    \centering
    \includegraphics[width=0.8\linewidth]{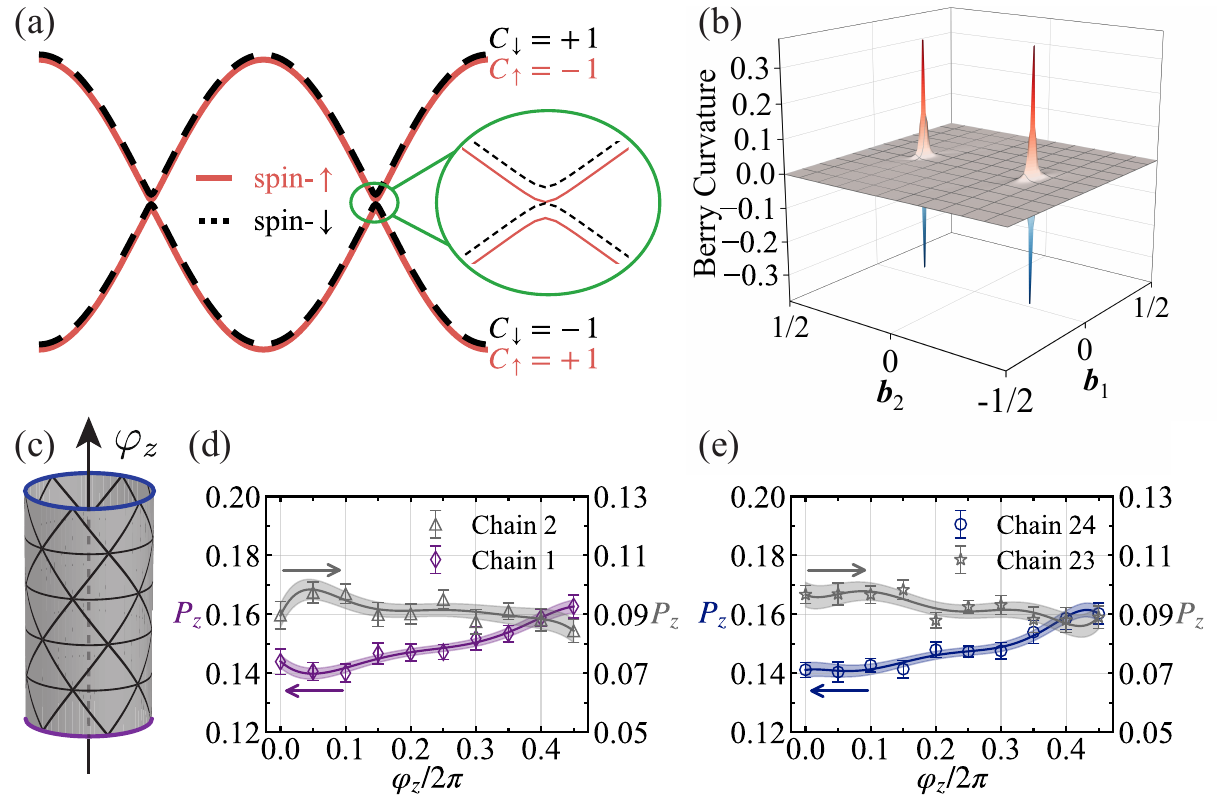}
    \caption{(a) Mean-field spinon dispersion for total magnetization $S^z_{\mathrm{tot}}=2$. The spin-$\downarrow$ valence band and spin-$\uparrow$ conduction band intersect at momenta $\pm \boldsymbol{b}_2/4$. (b) Berry curvature distributions for the spin-$\uparrow$ valence band (positive Berry curvature) and conduction band (negative Berry curvature).  The different sign between spin-$\uparrow$ and spin-$\downarrow$ spinon bands are due to time-reversal symmetry. (c) Schematic of a triangular lattice with cylindrical geometry. $\varphi_z$  represents the twist gauge flux. (d) and (e) The spin polarization $P_z$ near the bottom (Chain 1 and Chain 2) and top (Chain 23 and Chain 24) edges as a function of $\varphi_z$, respectively, with total magnetization $S^z_{\mathrm{tot}} = 2$. The system size in \textbf{a} and \textbf{b} is $240 \times 120$, while a $24\times24$ size is used in (d) and (e).}
    \label{fig:spin_hall}
\end{figure}
\section{Probing the Edge via Magnetic Field}

While the bulk is fully gapped and the chiral central charge vanishes due to time-reversal symmetry, the nontrivial topology of this SPT state is expected to manifest at the boundary. To access these protected edge properties, we introduce a magnetic field $B_{z}$  as a probe.
%Since the chiral central charge vanishes in this SPT state with time-reversal symmetry, we introduce a magnetic field $B_{z}$ to explore the edge properties. 
This involves adding the term $-B_{z}\sum_{i} S^{z}_{i}$ to the Hamiltonian (\ref{eq:model}). For the fermionic doublet, the additional term $-\sum_{i}\psi^{\dagger}_{i} \mu_{z} \sigma^{z} \psi_{i}$, resulting from the field $B_{z}$, is included in the mean-field Hamiltonian $H_{\mathrm{mf}}$ in equation~(\ref{eq:hmf}). To avoid disrupting the ground state, we aim to apply the weakest possible magnetic field. We found that the minimum system magnetization is $S^{z}_{\mathrm{tot}} = 2$ (see SM for details~\cite{supplement}). This indicates that the top of the mean-field valence band with spin down touches the bottom of the conduction band with spin up at two momenta $\pm \boldsymbol{b}_{2}/4$, as shown in Fig.~\ref{fig:spin_hall}(a). Additionally, Fig.~\ref{fig:spin_hall}(b) shows that the Berry curvature distribution is primarily concentrated near these contact points.

We then examine the spin pump at the magnetization of $S^{z}_{\mathrm{tot}} = 2$. Following Laughlin's argument~\cite{PhysRevB.23.5632}, we thread a flux $\varphi_{z} = \oint \boldsymbol{A}^{z}\cdot d\boldsymbol{l}$ through the cylindrical system [see the schematic diagram in Fig.~\ref{fig:spin_hall}(c)], so that a spinon acquires a phase of $e^{i\sigma\varphi_{z}/2}$ (for a spin-1/2 system) as it travels around the cylinder. We measure the spin pump between the two edges using the Gutzwiller-projected state $|\Psi(p_{\mathrm{opt}},\varphi_{z})\rangle$, as depicted in Fig.~\ref{fig:spin_hall}(d) and (e), where $P_{z}$ is defined as $P_{z} = \langle \Psi(\varphi_{z})|S^{z}_{\mathrm{chain}} |\Psi(\varphi_{z})\rangle/\langle \Psi(\varphi_{z})|\Psi(\varphi_{z}) \rangle$ with $S^{z}_{\mathrm{chain}} = \sum_{j \in \mathrm{chain}} S^{z}_{j}$ along a chain around the $z$-axis. The optimized parameters $p_{\mathrm{opt}}$ are fixed, so we omit them in the following for simplicity. The non-zero slopes of the curves indicate that the spin Hall conductivity is finite but not quantized, unlike in CSL~\cite{PhysRevLett.127.087201}. Notably, the spin polarization curves of adjacent chains consistently show opposite slopes, while their sum remains approximately constant within numerical error. This behavior occurs not only near the edges (chains 1 and 2 at the bottom edge, and chains 23 and 24 at the top edge) but also within the bulk (see SM for details~\cite{supplement}), indicating a standing-wave pattern with a characteristic phase approximately $\boldsymbol{G}_{v} \sim \boldsymbol{b}_2/2$. To capture this phenomenon quantitatively, we define an interchain structure factor:
\begin{equation}
    P_{z}(\varphi_{z},\boldsymbol{k}) = \sum_{n,m}e^{i\boldsymbol{k} \cdot (\boldsymbol{r}_{n} - \boldsymbol{r}_{m})} \langle S^{z}_{n}S^{z}_{m} \rangle_{\varphi_{z}},
\end{equation}
where $n$ and $m$ are chain indices, $\boldsymbol{r}_{n(m)} = n(m)\boldsymbol{a}_2$ represents the location of the $n$($m$)-th chain, and $\langle S^{z}_{n}S^{z}_{m} \rangle_{\varphi_{z}} = \langle \Psi(\varphi_{z})|S^{z}_{n}S^{z}_{m} |\Psi(\varphi_{z})\rangle/\langle \Psi(\varphi_{z})|\Psi(\varphi_{z}) \rangle$. The structure factor reaches an extremum near $\boldsymbol{G}_{v}$ (see SM for details~\cite{supplement}), reflecting the underlying standing-wave pattern. This phase is closely related to the Berry curvature distribution [see Fig.~\ref{fig:spin_hall}(d)] concentrated near two valley momenta $\boldsymbol{b}_2/4$, giving rise to valley interference $\boldsymbol{G}_{v} = \boldsymbol{b}_2/4 - (-\boldsymbol{b}_2/4)$. To verify this, we artificially adjust the parameter $t_{z,ij}$ to make the Berry curvature distribution more uniform. Under this modification, the standing-wave behavior weakens significantly (see SM for details~\cite{supplement}). 

The alternating spin-Hall response uncovered here represents a novel form of bulk-edge correspondence tailored to this confined SPT phase: while the bulk is rendered featureless by Polyakov confinement, the Berry curvature inherited from the parent Dirac spin liquid resurfaces as a real-space standing wave along the open boundary. This valley-interference pattern—encoded in the chain-resolved spin texture under a weak magnetic field—provides both a diagnostic of the underlying topology and a concrete observable for experimental detection.

%These observations indicate that the alternating spin-Hall response between adjacent chains, with nearly opposite slopes and a constant total across each pair, originates from the concentration of Berry curvature near the valley momenta. This concentration creates a standing-wave pattern along the open-boundary direction, effectively forming a spin-flow texture across chains. This chain-resolved texture directly reflects the valley interference encoded in the Berry curvature distribution and provides a clear signature of how momentum-space Berry curvature translates into real-space spin transport patterns.

\section{Summary}

In summary, we construct a time-reversal-symmetric SPT state within the Polyakov confinement framework by coupling a spin-dependent gauge field to the spinons of the Dirac spin liquid, which exhibits a quantum spin Hall–type response. We introduce a loop-based diagnostic of confinement by evaluating the scaling behavior of the string-induced loop quantity $W[\mathcal{C}]$. Its distinct dependence on the enclosed area and perimeter provides direct numerical evidence for spinon confinement in the bulk and clearly distinguishes the SPT state from the deconfined DSL. We determine the phase diagram of the antiferromagnetic Heisenberg model on the triangular lattice using VMC calculations, demonstrating that the interplay between the next-nearest-neighbor coupling $J_2$ and the DM interaction leads to the realization of this SPT state. Complementarily, we extract the TEE using the Kitaev–Preskill construction and find it to vanish within numerical accuracy, consistent with the short-range-entangled nature of the bulk state. Additionally, we investigate the spin-pump response of the SPT state in the presence of an external magnetic field, uncovering characteristic edge transport signatures. These responses encode the Berry-curvature structure around the original Dirac spinon cones and offer experimentally accessible probes to identify this confined SPT state. More broadly, our work establishes a concrete route to realizing interaction-driven SPT phases through gauge-field confinement and bridges field-theoretic confinement mechanisms with microscopic numerical diagnostics in quantum spin systems. Our results may be testable in future experiments or quantum simulation platforms~\cite{nat.645.341}.

%
% Each of the commands below will create an unnumbered section with the appropriate heading.
% Remove any sections that are not relevant for your article.
% All sections except suppdata will be removed if the [anonymous] option is used.
% See iopjournal-guidelines.pdf for more information.
%

% \section{Supplementary information}
% See Supplementary information for detailed information on the Chern-Simons effective action, topological responses, and calculation details in VMC.

\ack{This work was supported by National Key Projects for Research and Development of China (No. 2021YFA1400400 and No. 2024YFA1408104) and the National Natural Science Foundation of China (No. 12547122, No. 12374137, No. 12434005, and No. 12550405).}

% \funding{Sample text inserted for demonstration.}
% This section is a list of funder names and grant numbers

\data{The data that support the findings of this study are available from the authors upon reasonable request.}

Supplementary material available at https://doi.org/xxx.
% For more information on IOP Publishing's research data policy see: https://publishingsupport.iopscience.iop.org/questions/research-data/

\roles{
Li-Wei He\orcid{0009-0001-9897-545X}\\
Conceptualization (equal), Investigation (equal), Writing – review \& editing (equal)\\

Shun-Li Yu\orcid{0000-0001-7202-4851}\\
Conceptualization (equal), Investigation (equal), Writing – review \& editing (equal)\\

Jian-Xin Li\orcid{0009-0007-3901-8119}\\
Conceptualization (equal), Writing – review \& editing (equal)\\}
% List author names and the contributions made to the article, using terms from the NISO Contributor Roles Taxonomy (CRediT) https://credit.niso.org

% \suppdata{See Supplementary material~\cite{supplement} for detailed information on the Chern-Simons effective action, topological responses, and calculation details in VMC.}

\bibliographystyle{iopart-num}
\bibliography{ref}

\end{document}